\documentclass[letterpaper,twocolumn,10pt]{article}
\usepackage{usenix,epsfig,endnotes}
\begin{document}

\date{}

\title{\Large \bf Towards Resilience and Autonomy-based Approaches for Adolescents Online Safety}

\author{
{\rm Jinkyung Park}\\
Vanderbilt University
\and
{\rm Mamtaj Akter}\\
Vanderbilt University
\and
{\rm Naima Samreen Ali}\\
Vanderbilt University
\and
{\rm Zainab Agha}\\
Vanderbilt University
\and
{\rm Ashwaq Alsoubai}\\
Vanderbilt University
\and
{\rm Pamela Wisniewski}\\
Vanderbilt University}

\maketitle

\thispagestyle{empty}

\subsection*{Abstract}
In this position paper, we discuss the paradigm shift that has emerged in the literature, suggesting to move away from restrictive and authoritarian parental mediation approaches to move toward resilient-based and privacy-preserving solutions to promote adolescents' online safety. We highlight the limitations of restrictive mediation strategies, which often induce a trade-off between teens' privacy and online safety, and call for more teen-centric frameworks that can empower teens to self-regulate while using the technology in meaningful ways. We also present an overview of empirical studies that conceptualized and examined resilience-based approaches to promoting the digital well-being of teens in a way to empower teens to be more resilient.


\section{Introduction}
According to a Pew Research report, 97\% of U.S. teens use the internet daily; 46\% of them are online almost constantly. Most teens have access to digital devices such as smartphones (95\%), desktop or laptop computers (90\%), and use at least one social media platform (95\%) ~\cite{vogels_2022_pewresearch}. While the majority of U.S. teens shared that being on social media provided them a space for social connection, creativity, and peer support~\cite{Anderson_2022_pewresearch}, recent research has shown evidence that teens experience many online risks, such as cyberbullying~\cite{kim2021human}, exposure to explicit content~\cite{park2023towards}, problematic internet use with mental health problems (e.g., suicide challenges)~\cite{park2023affordances}, and even physical safety risks~\cite{valkenburg2022social}. Reactions to these new risk experiences were an overemphasis on restrictive and authoritative parental mediation practice~\cite{modecki2022}. The use of restriction and monitoring by parents may shield teens from online risks, but at the cost of trust between parents and teens, and positive family value as a whole~\cite{rutkowski2021family, wisniewski2017parental}. 
In this position paper, we discuss the paradigm shift that has emerged in the literature, moving away from restriction and privacy-invasive parental mediation toward resilient-based and privacy-preserving interventions to promote teen online safety. We then provide an overview of empirical studies that our research teams are working on to conceptualize resilience-based approaches to promoting the digital well-being of teens in a way to empower teens to be more resilient. Our position paper is highly relevant to KOPS 2023 as our focus on resilience-based approaches to teen online safety is one of the core topics of interest for KOPS 2023 (i.e., resilience factors associated with minors' crime and safety). 

\section{Background}
In this section, we synthesize existing literature that reflects a paradigm shift from emphasizing restrictive monitoring to promoting resilient-based and privacy-preserving approaches to promote teen online safety.

\subsection{Restriction as a Mean for Protecting Teens from Online Risk}
Past literature on parental mediation and adolescent online safety often puts a strong emphasis on abstinence-based strategies (i.e., parental control app) to restrict and monitor teens’ online activities~\cite{ schiano2017parental, wisniewski2017parental, khurana2015protective}. 
Through the parental control apps, fine-grained details about teens’ smartphone use, such as websites visited, apps installed, calls made, texts sent (including the actual content of the message), and geo-location, are often shared with parents~\cite{wisniewski2017parental}. The rationale behind such tools is to make sure that teens are engaging in age-appropriate ways online and are being sufficiently monitored by their parents when doing so. 
However, the use of restrictive approaches to monitor their mobile online activities comes with the cost of teens’ autonomy~\cite{baumrind2005patterns}, trust between parents and teens~\cite{williams2003adolescents}, and positive family value as a whole~\cite{wisniewski2017parental}. More critically, teens such as foster youth are more vulnerable to online risks as they often do not have parents to actively engage with them to ensure their online safety~\cite{badillo2017understanding, badillo2017abandoned}. 

Social media platforms have also taken restrictive measures to protect youth for instance, by preventing adults from sending private messages to minors they are not connected using sensitivity filters and advanced parental controls ~\cite{instagram2021restrictingdms, instagram_2021_contentcontrol}. Implementation of such safeguards by social media platforms is grounded by the U.S. federal law, the Children’s Online Privacy Protection Act (COPPA)~\cite{coppa2013}, which provided a strong legal ground for the big-tech companies to protect children under 13. Such safeguards, however, do not apply to youth over 13, who have been largely left out of broader public policy debates and self-regulatory industry programs~\cite{montgomery2015youth}. 


\subsection{Trade-offs Between Privacy and Protection}

While teens place great social value on their online privacy and appreciate rules and policies that are fair and negotiable ~\cite{steeves2014young}, as designers and parents, we develop and use surveillance technologies that take teens’ privacy away for the sake of their online safety. Scholars, policymakers, security professionals, and advocates have continuously discussed the effects of surveillance/monitoring technologies on individuals and society altogether~\cite{vcas2017introduction}. In part, complexities and controversies around privacy vs protection discourse are due to a lack of legal frameworks within the U.S. balancing teen online safety and protecting teen privacy rights. Recently, a comprehensive bipartisan legislation, the Kids Online Safety Act (KOSA), has been proposed by U.S. Senators~\cite{Blumenthal2022}. The legislation requires social media platforms to proactively mitigate harm to minors such as the promotion of self-harm, suicide, and sexual exploitation. It also requires independent audits and public scrutiny from experts to ensure that parents and policymakers know whether social media platforms are taking meaningful steps to address risks to kids~\cite{Blumenthal2022}. 

Yet, controversies have continued to rise around KOSA. For instance, teen privacy advocates fear that this legislation would incentivize social media sites to collect \textit{even more information about children} to prevent a set of harms to minors. More importantly, the advocates argued that KOSA could effectively be an instruction for social media platforms to employ a broad range of content filtering to limit minors’ access to certain online content, such as sex education for LGBTQ+ youth (which schools had implemented in response to earlier legislation)~\cite{Feiner2022}. 
Given the repeated failures of social media companies to protect youth from serious risks on their platforms~\cite{Wired, CNN, WSJ}, the idea of developing legal frameworks to monitor youth online risk is promising, but only with additional legal safeguards to respect teens' privacy in place. Such legal frameworks are necessary, especially when some of the technical solutions (e.g., AI-based risk detection) rely heavily on teens' intimate data. At the same time, there should be more teen-centric and privacy-preserving solutions that empower teens to self-regulate their online risk experiences and thrive as healthy digital citizens. 


\subsection{Empowering Teens through Resilience-based Interventions and Co-Design with Teens}

Teens need autonomy to individuate themselves from their parents, but at the same time, they are less capable than adults at managing online risks without guidance~\cite{wisniewski2022privacy}. Hence, there is a need for strength-based design practices that can empower teens and parents to manage online risks in meaningful ways. The key idea around this approach is the paradigm shift from an authoritarian view of protecting teens to more \textbf{supportive frameworks that can empower teens to self-regulate and manage technology use in meaningful ways}~\cite{ghosh2018matter, wisniewski2022privacy}. In academia, there has been a push for these new approaches respecting teens' rights in the online space, building resilience, and active involvement of both parents and teens to manage teens' online life. 

One line of research has conceptualized more collaborative technologies that move away from surveillance-based approaches to ones that engage both teens and parents for digital rule-setting and managing online activities~\cite{akter2022parental,ghosh2018safety,nouwen2017parental, ko2015familync}. For instance, Ghosh et al.~\cite{ghosh_circle_2020} developed a mobile app that allows adolescents and parents together to negotiate trusted vs. untrusted contact to exchange text messages and confirmed that parents and adolescents who appreciated family values such as privacy, trust, freedom, and balance of power preferred the app over the traditional parental control apps. Another way to support teens' resilience and autonomy is through participatory design that puts teens as the primary stakeholder and authority of their own online experiences. Co-design research with teens has been successful in including teen voices and their unique perspectives to design resilience-based approaches for promoting their online safety~\cite{agha2022case,razi2020let, ashktorab2016designing}. By giving voice to the design process of their online safety solutions, teens can reflect on their own online habits and learn how to self-regulate such habits in ways that promote resilience, autonomy, and digital well-being~\cite{chatlani2023teen}. 

\section{Research on Resilience-based Approaches to Support Teen Online Safety}
Over the past decade, our research team has been working on several research projects related to the topic of adolescents' online safety and risks \cite{agha2022case, agha2023strike, akter_coops_2022, akter2022parental, akter_it_2023, badillo2019risk, badillo2019stranger, badillo2017abandoned, chatlani2023teen, ghosh_circle_2020, ghosh_safety_2018, ghosh2018safety, ghosh2018matter, razi2021human, razi2020let, razi2020deploying, alsoubai2023human, alsoubai2022friends, alsoubai2022mosafely}. In this section, we will summarize how our ongoing research moves beyond
traditional approaches relying on unidirectional restrictive and privacy-invasive mechanisms, toward resiliency and autonomy-based design that can
empower teens to utilize their knowledge to self-regulate and cope in the face of online risks.

\subsection{Youth Advisory Board/Teenovate}

Teen-centered Participatory Design (PD) programs have made successful efforts in involving teens directly in the design process of online safety interventions \cite{chatlani2023teen, agha2022case}. Still, the current PD practices are found to be insufficient in retaining teen participants who may not find the program rewarding enough, or perceive power imbalances between themselves and adult researchers pertaining to knowledge gaps \cite{poole2013interaction, davis2020co}. Therefore, the authors engaged 21 teens to plan and develop Teenovate Youth Advisory Board (YAB) \cite{chatlani2023teen}, a longitudinal participatory action research program for teens.  
In YAB, teens act as co-researchers to give feedback on the proposed adolescent online safety research and do design activities to promote their own online safety. Through the participatory sessions, YAB teens provided us with insights such as how to improve the working relationships with teen participants or behavioral approaches for resolving possible conflicts between teens and adult researchers. Additionally, YAB offers training workshops focused on human-centered research and User Experience (UX), to equip teens with the necessary tools to make strides and effectively contribute to designing various solutions for their safety online. 

With the YAB members, the authors have also explored the potential of the Asynchronous Research Community (ARC) \cite{macleod2016asynchronous} methodology to improve engagement by encouraging them to post discussion comments and interact with peer participants and researchers on topics related to adolescent online safety research on Discord~\cite{jean2023teens}. Through the ARC study, the authors found that teens are most interested in exploring topics related to their online privacy on social media. YAB teens shared that they consciously decide which social media platforms to use depending on their user goals (e.g., content viewing or socializing), and employ varied strategies to strike a balance between meeting their goals and mitigating privacy risks. The findings suggest that understanding teens' motivations and needs for social media use and their differing privacy perceptions is pivotal to providing them with customized support systems for safer online experiences~\cite{jean2023teens}. 

One of the key lessons learned from the YAB teens was that teens consider the co-design sessions as an opportunity to channel their creativity skills and online experiential knowledge to devise tangible solutions for their online safety. They also found benefits of participating in adolescent online safety research as it allowed them to reflect on their own online interactions and reconsider the online safety measures that would work best for them. Overall, participating in co-design activities helped teens feel more confident and empowered to contribute freely to the research aimed at promoting their online safety. In the future, these insights will serve as the guiding beacon to generate study design patterns for researchers suggesting best practices to conduct online safety research for teens with teens.

\subsection{Joint Family Oversight for Adolescents' Mobile Online Safety, Security and Privacy}
In families, parents often use parental monitoring or controlling apps to ensure their teens' mobile online safety. However, teens often find this constant surveillance overly restrictive and privacy-invasive, affecting parent-teen relationships \cite{wisniewski_parental_2017, ghosh_safety_2018}. Therefore, adolescents online safety researchers have called for adopting more collaborative \cite{cranor_parents_2014, hashish_involving_2014, kropczynski_examining_2021} and teen-centric approaches where teens can have some level of privacy and autonomy in their own online safety \cite{charalambous_privacy-preserving_2020, ghosh_circle_2020}. In an attempt to move toward more bi-directional approaches for mobile online safety, Akter et al. ~\cite{akter2022parental} leveraged the concepts of the community oversight model for privacy and security \cite{chouhan_co-designing_2019} and developed a joint family oversight mechanism, titled CO-oPS \cite{akter_coops_2022, akter_evaluating_2023, akter2024familydesign}, to help parents and teens collaboratively manage their mobile privacy, online safety, and security. This CO-oPS app provided a transparent view of one another's apps installed and the privacy permissions granted or denied to the apps and allowed to provide direct feedback to one another, facilitating more parent-teen communication. However, it also allows users to manage the level of transparency by allowing them to hide any of the apps installed that they do not want to be shared with other users.

Through a lab-based study with 19 parent-teen dyads, Akter et al. ~\cite{akter2022parental} evaluated the CO-oPS app to assess whether it would be useful for their families in managing mobile online safety, security, and privacy. They found that even though teens were the primary tech support providers in the family, parents were concerned for their teens’ mobile online safety and, therefore, often manually checked their phones to review the apps installed or used parental control apps to restrict new app installation. In evaluating the CO-oPS app, both parents and teens found value in the feature that allowed them to review one another’s apps and permissions, as it increased the transparency of their app usage and helped initiate more discussion around mobile privacy and security. The authors also observed that parents were more concerned about teens’ app usage, seeing them as access points to their children by others online. In contrast, teens mostly focused on the permissions granted on their parent's phones, being more aware of the malicious intent of third-party mobile apps. However, they also found that such bidirectional co-monitoring made parents uncomfortable relinquishing control over their teens. In contrast, teens felt it was not their place to oversee their parents' mobile privacy and security.

Overall, Akter et al. \cite{akter2022parental} demonstrated that by encouraging more collaborative monitoring and communication, bidirectional oversight between parents and adolescents might potentially increase families' overall knowledge and awareness of mobile privacy, security, and online safety, which they further confirmed in a later study \cite{akter_it_2023}. However, for these beneficial results to take place, both parents and adolescents must agree that it is their responsibility to look out for one another, which requires a significant paradigm shift from the adolescent online safety approaches that have been predominantly used in our society. Additionally, the insights derived from employing CO-oPS within family contexts \cite{akter2022parental} can offer advantages not only to designers developing similar joint family online safety tools but also to designers creating strategies aimed at assisting users in safeguarding their digital privacy. This extends beyond the domain of mobile online safety to encompass broader areas of digital privacy and security, including smart home devices \cite{alghamdi_codesigning_2023, alghamdi_misu_2022, emami_influence_2018}, social media \cite{bhagavatula_adulthood_2022, such_privacy_2016, vishwamitra_towards_2017, ulusoy_panola_2021}, websites \cite{mcdonald_citizens_2021, mcdonald_building_2021}, and other platforms where teenagers and their families share personal information with third parties.

\subsection{Youth-centered Online Risk Detection}
Computational approaches to identify teen online risk have been applied as promising alternatives for human labor to do the same tasks, given the scale of online content. Yet, a common trend among these approaches is a lack of teen-centered aspects~\cite{razi2021human, kim2021human, alsoubai2022mosafely}, such as a lack of understanding of the context of risks teens experience~\cite{alsoubai2022friends}. In this study, the focus was to improve our understanding of teens’ online risk experiences across online and offline contexts by aligning their self-reported surveys with explicit risk flagging, the evidence of which can help provide youth more agency for their online/offline safety. In this IRB-approved user study, the authors worked with 173 youths (ages 13-21) to collect self-reported surveys regarding their online and offline risk experiences. Then, youths  were asked to upload their private social media (i.e., Instagram Direct Messages) data to self-assess the risks in their private conversations (e.g., risk type, level of risk, etc.)~\cite{razi2022instagram}. The authors created profiles of youth based on their self-reported survey data and compared the profiles with the risk type and levels that the youth flagged. Five unique profiles of youth emerged: 1) Low Risks, 2) Medium Risks, 3) Increased Sexting, 4) Increased Self-Harm, and 5) High-Risk Perpetration. 

The comparative analysis confirmed that youth self-reported online and offline risk experiences were fairly aligned with their social media trace data and that youth risk experiences varied depending on their profiles. For instance, while youths in the low-Risk group were exposed to spam and scam messages and the self-harm disclosures of others, those in the medium-risk group mostly encountered harassment. Another key finding was the mismatch between offline and online risk contexts. For example, youth in the Increased Self-Harm profile group reported the highest levels of offline self-harm, but their unsafe conversations did not contain digital self-harm content; instead, they engaged in more unsafe sexual conversations. The results of the study highlighted the importance of understanding the multidimensionality of youth online risk experience as it is pivotal for designing youth-centric and customized risk prevention strategies to promote youth resilience from online risk. More importantly, having such in-depth knowledge of youth online risk experience is critical to inform the design and development of youth-centered computational systems to identify nuanced and contextualized online harms that youth experience at scale. Based on the knowledge gained from this study, the authors are working on designing youth-centered “real-time” risk detection models as ‘just-in-time’ interventions to mitigate their online risk experience. 

\subsection{Real-time Nudge-based Intervention}
“Nudges” are subtle cues that aim to influence people’s behavior without compromising their decision-making autonomy~\cite{thaler2009nudge} and have been proposed as a ‘just-in-time’ intervention to support teens at the moment when they experience risks online ~\cite{agha2022case, agha2023strike}. Yet, a majority of the prior work within the online safety space has focused more on \textit{designing} interventions ~\cite{badillo2019stranger, hartikainen2019children}, with less realistic evaluations to assess the effectiveness nudges for teen online safety. Moreover, the few evaluations of adolescent online safety interventions have relied on self-reported survey-based feedback \cite{masaki2020exploring}, which is subject to recall bias and does not provide an ecologically valid setting for evaluations. To overcome this, one way for evaluating nudges is by simulating the environment and risks to understand how nudges lead to actual behavior change for adolescent online safety. Such simulation-based evaluations have been studied as a promising approach within other related fields of privacy and security ~\cite{zinkus2019fakesbook, kumaraguru2010teaching}. In order to design realistic bad actors and risk scenarios for nudge evaluation within a Social Media Simulation, it is crucial to involve teens in designing such a simulated environment. 

Therefore, the authors conducted co-design sessions with 14 teens (13-18 years old) in the United States to obtain their feedback on the design of user personas and risk scenarios which will be implemented in a social media simulation, for evaluating adolescents' online safety interventions~\cite{agha2023co}. During these sessions, the authors presented teens with 10 prepared user personas and 4 risky scenarios based on prior research~\cite{agha2023strike}. Teens redesigned various aspects of the social media personas and scenarios using an online whiteboard tool, FigJam~\cite{Figjam}. The results show that teens considered the characteristics of the risky user to be important and designed personas to have traits that align with the risk type, were more believable and authentic, and attracted teens through materialistic content. Teens also redesigned the risky scenarios to be subtle in information breaching, harsher in cyberbullying, and convincing in tricking the teen. The findings from the co-design sessions emphasized the importance of designing simulations that are sensitive to the needs and perspectives of teens and that provide a nuanced and realistic environment for evaluating online safety interventions~\cite{agha2023co}. Moving forward, the authors are planning to implement the designs from this study in a between-subjects experimental design with teens to evaluate the effectiveness of the different types of nudges within a social media simulation. 

\section{Conclusion}
In this position paper, we highlighted a paradigm shift that has emerged in the literature in a way that moves away from restriction and authoritarian parental mediation toward resilient-based and privacy-preserving solutions to promote teen online safety. We also provided an overview of empirical studies that conceptualized and examined various approaches to promoting the digital well-being of teens in a way to empower teens to be more resilient. A common theme among the studies that we introduced above is a call for teen-centered approaches to promoting teens' digital well-being while supporting the healthy development of teens. As many of the studies are ongoing projects, more empirical evidence of the benefits of resilience-based interventions will be documented. Having said that, participating in KOPS 2023 would be extremely beneficial for us to share ongoing projects and have interactive feedback from the workshop participants, which could potentially lead to more collaboration opportunities. Finally, we anticipate learning more about others' translational research to promote teen resilience by participating in the workshop.

\bibliographystyle{plain}
\bibliography{Reference}

\end{document}